\documentclass[12pt]{article}

\usepackage{amsmath}
\usepackage{graphicx}
\usepackage{amsfonts}
\usepackage{amssymb}
\usepackage{epsfig}
\usepackage{color}
\usepackage{psfrag}

\begin{document}
\setcounter{page}{0} \topmargin 0pt \oddsidemargin 5mm
\renewcommand{\thefootnote}{\arabic{footnote}}
\newpage
\setcounter{page}{0}

\begin{titlepage}

\begin{flushright}
ISAS/63/2003/FM\\
%ITF XXX
\end{flushright}
\vspace{0.5cm}
\begin{center}
{\Large {\bf Finite--Volume Form Factors in }}\\
{\Large {\bf Semiclassical Approximation}} \\

\vspace{2cm}
{\large G. Mussardo$^{1,2}$, V. Riva$^{1,2}$ and G. Sotkov$^{3}$} \\
\vspace{0.5cm} {\em $^{1}$International School for Advanced Studies}\\
{\em Via Beirut 1, 34100 Trieste, Italy} \\
\vspace{0.3cm} {\em $^{2}$Istituto Nazionale di Fisica Nucleare}\\
{\em Sezione di Trieste}\\\vspace{0.3cm} {\em $^3$ Instituto de Fisica Teorica} \\
{\em Universidade Estadual Paulista} \\
{\em Rua Pamplona 145, 01405-900 Sao Paulo, Brazil}

\end{center}
\vspace{1cm}

\begin{abstract}
\noindent
A semiclassical approach is used to obtain Lorentz covariant expressions
for the form factors between the kink states of a quantum field theory
with degenerate vacua. Implemented on a cylinder geometry it provides
an estimate of the spectral representation of correlation functions in
a finite volume. Illustrative examples of the applicability of the method
are provided by the Sine-Gordon and the broken $\phi^4$ theories
in 1+1 dimensions.

\end{abstract}

\end{titlepage}

\newpage

\section{Introduction}\label{intro}

The problem of understanding quantum field theories (QFT) on a
finite volume is of great importance both for theoretical
interests and for many realistic applications. In particular, it
is crucial to explore the possibility of defining a "form factor
representation" for the correlation functions, in analogy to the
infinite volume case \cite{ffstructure}. There are reasons to
expect, in fact, a fast convergent behaviour of these series also
for finite volume correlators, as it happens in infinite volume
(see, for instance \cite{on}). If this would be indeed the case,
accurate estimates of finite volume correlators and other related
physical quantities, could be obtained by just using few exact terms
of their spectral representations, having consequently a great simplification
of the problem. This observation makes clear that it is worth pursuing
the research on this topic and looking, in particular, for an efficient
scheme of approximation.

It is well known that finite volume quantum field theories are
also related to field theories at finite temperature. In 1+1
space-time dimensions, this gives the possibility of viewing a QFT
defined on a finite volume of a cylinder geometry in the
alternative picture, in which the space variable is infinite and
the (euclidean) time variable is instead compactified on a circle
of radius $R=1/T$, where $T$ is the temperature of the system (the
so-called Matsubara imaginary time formalism). In the case of
integrable theories, promising exact results have been obtained in
\cite{lecmuss}, thanks to the Thermodynamic Bethe Ansatz (TBA)
technique, which relates the thermodynamical properties of the
system to its exact $S$-matrix \cite{ZamBethe}. As shown in
\cite{lecmuss}, the finite temperature correlation functions can
be expressed in terms of the usual form factors, relative to the
infinite volume Hilbert space, dressed with the so-called
"filling-fractions", which encode the thermodynamics of the
system. Within this approach, it has been possible to fully
understand the finite temperature one-point functions, while a
further analysis seems still necessary for the two-point and
higher correlation functions.

Coming back to the picture where the time evolution is the usual
one while the space variable is compactified, the form factors
have to be seen as the matrix elements of local fields between
eigenstates of the finite volume hamiltonian. In this case there
is still a poor understanding of the correlation functions
behaviour, even in the integrable cases. It must be stressed that
the reason is not only the present undeveloped and unsatisfactory
theory for computing finite volume matrix elements. In fact,
contrary to the infinite volume situation, the only knowledge of
the form factors is not enough to obtain the correlators in a
finite volume since for their computation the energy eigenvalues
of the finite volume hamiltonian are also needed. For integrable
models, these extra data may be obtained along the lines discussed
in \cite{bazh}, while for non--integrable theories one has necessarily
to rely on numerical methods, as the one proposed in \cite{trunc} for
instance.

Leaving apart the above problem of energy eigenvalues and concerning
instead about the status of the form factors in a finite volume, there
are so far only semiclassical results relative to a conformal theory
\cite{fsmirnov} as well as exact calculations relative to the Ising model
\cite{fonszam} (for a Bethe Ansatz approach see, however, \cite{korepin}).
Although these findings are very interesting, the techniques employed in
the above papers are however strictly related either to the
specific integrable structure of the considered models or to the
free nature of the Majorana fermion field of the Ising model. In
this paper we are going to use, instead, a semiclassical approach
which does not require integrability and it may be then of more
general applicability, of course within its range of
approximation. As shown below, this approach gives in particular
the possibility of checking some of the results of our analysis
against the exact known quantities in the case of integrable
theories in infinite volume, while it permits to have new
predictions on the semiclassical regime of the non--integrable
ones. As illustrative examples of this method, we will present its
application to two significant models: the integrable QFT of the
Sine-Gordon model
\begin{equation}\label{SGpot}
V(\phi)=\frac{m^{2}}{\beta^{2}}(1-\cos\beta\phi) \,\,\,,
\end{equation}
and the non--integrable QFT with a $\phi^{4}$ interaction
in the broken $Z_2$ symmetry phase
\begin{equation}\label{phi4pot}
V(\phi) = \frac{\lambda}{4}\phi^{4}-\frac{m^{2}}{2}\phi^{2}
+ \frac{m^{4}}{4\lambda}\;.
\end{equation}
Both these theories display degenerate minima, that can be chosen
as the constant classical solution representing the vacuum state
around which one decides to quantize the theory. However, the most
interesting classical backgrounds are the kink--type ones, which
interpolate between degenerate minima and give rise to a non
trivial quantization scheme. Our aim is to define this kind of
backgrounds in finite volume and to estimate the relative form
factors in an appropriate limit. At the same time, doing so, we
will have the possibility of better exploring the properties of
the same form factors in the infinite volume case.

\section{Classical solutions and form factors}\label{generalsect}

The semiclassical quantization of a field theory defined by a
potential $V(\phi)$ is based on the identification of a classical
background $\phi_{cl}(x)$ which satisfies the equation of motion
\begin{equation} \label{eom}
\partial_{\mu}\partial^{\mu}\phi_{cl}+V'(\phi_{cl})=0\;.
\end{equation}
In the infinite volume, this can be performed with various well
established techniques, like the path integral formalism
\cite{pathint} or the solution of the field equations in classical
background \cite{dashen}, usually called the DHN method (for a
systematic review, see \cite{raj}).

The procedure is particularly simple and interesting if one
considers classical field solutions $\phi_{cl}(x)$ in 1+1
dimensions which are static "kink" configurations interpolating
between degenerate minima of the potential, and whose quantization
gives rise to a particle-like spectrum. The kink solutions are
obtained, firstly, by integrating the first order equation related
to (\ref{eom})
\begin{equation}
\label{first}
\frac{1}{2}\left(\frac{\partial \phi_{cl}}{\partial
x}\right)^{2} = V(\phi_{cl}) + C \,\,\,,
\end{equation}
and, secondly, imposing that $\phi_{cl}(x)$ reaches two different
minima of the potential as $x\rightarrow\pm\infty$. In the
infinite volume case, these boundary conditions correspond to
fixing the constant of integration $C$ equal to zero. As we will
show below, finite volume case can be described, on the contrary,
by a non--vanishing value of $C$ which can be directly related to
the size of the system. In the following we will be interested to
discuss QFT defined on a cylinder geometry where the time variable
is infinite while the space one is compactified on a circle of
circumference $L$, with twisted boundary conditions, i.e.
\begin{eqnarray*}
\phi(x+L)&=&2\pi-\phi(x)\qquad \textrm{for Sine-Gordon}\\
\phi(x+L)&=&-\phi(x) \qquad \quad\;\;\textrm{for $\phi^{4}$
theory}
\end{eqnarray*}

\vspace{0.5cm}

A very interesting result, due to Goldstone and Jackiw
\cite{goldstone} and well explained in \cite{raj}, is that the
classical background $\phi_{cl}(x)$ has the quantum meaning of
Fourier transform of the form factor of the basic field $\phi(x)$
between kink states\footnote{The classical solution can
also be directly seen as the matrix element of $\phi$ between
asymptotic states in the space coordinates representation:
%\\
$<x_{2}|\,\phi(0)|\,x_{1}>=\delta(x_{1}-x_{2})\;\phi_{cl}(x_{1})$.}.
The technique to derive this result relies
on the basic hypothesis that the kink momentum is very small
compared to its mass, which is indeed inversely proportional to
the coupling constant, considered small in the semiclassical
regime. Calling $p_{1}$ and $p_{2}$ the momenta of the in and out
one-kink states, at the leading order in the coupling constant one
obtains
\begin{equation}
<p_{2}|\,\phi(0)|\,p_{1}>=\int
da\;e^{i(p_{1}-p_{2})a}\;\phi_{cl}(a)\;.
\end{equation}
This important result has, unfortunately, two serious drawbacks:
expressing the form factor as a function of the difference of
momenta, Lorenz covariance is lost and moreover, the antisymmetry
under the interchange of momenta makes problematic any attempt to
go in the crossed channel and obtain the matrix element between
the vacuum and a kink--antikink state.

In order to overcome these problems, we need to refine the method
proposed in \cite{goldstone}. This can be done by using
the rapidity variable $\theta$ of the kink (and considering it as
very small), instead of the momentum. For example, in the $\phi^{4}$
theory (\ref{phi4pot}), where the kink mass $M$ is of order
$1/\lambda$, we will work under the hypothesis that $\theta$ is of
order $\lambda$. In this way we get consistently
\begin{equation}\label{rapidity}
E\equiv M \cosh\theta\simeq M\;,\qquad
p\equiv M \sinh\theta\simeq
M\,\theta \ll M\;.
\end{equation}
We can now define the form factor between kink states as the Fourier
transform with respect to the Lorenz invariant rapidity difference
$\theta\equiv\theta_{1}-\theta_{2}$:
\begin{equation}\label{ffinf}
<p_{1}|\phi(0)|p_{2}>\equiv f(\theta)\equiv M\int
da\,e^{i\,M\theta a}\hat{f}(a)\;,
\end{equation}
with the inverse Fourier transform defined as
\begin{equation}
\hat{f}(a)\equiv \int \frac{d\theta}{2\pi}\,e^{-i\,M\theta
a}f(\theta)\;.
\end{equation}
Following the same procedure used by Goldstone and Jackiw
\cite{goldstone}, i.e. starting from the Heisenberg equation of
motion for the field $\phi(x,t)$
\begin{equation}
\left(\partial_{t}^{2}-\partial_{x}^{2}\right)\phi(x,t)=-V'[\phi(x,t)]\;,
\end{equation}
and taking the matrix elements of both sides
\begin{equation}\label{heis}
\left[-(p_{1}-p_{2})_{\mu}(p_{1}-p_{2})^{\mu}\right]
e^{-i(p_{1}-p_{2})_{\mu}x^{\mu}}<p_{1}|\phi(0)|p_{2}> \,=\,
\end{equation}
\begin{equation*}
-e^{-i(p_{1}-p_{2})_{\mu}x^{\mu}}<p_{1}|V'[\phi(0)]|p_{2}>\;,
\end{equation*}
it is easy to show that, at leading order in $\lambda$, the function
$\hat{f}(a)$ obeys the same differential equation satisfied by the
kink solution, i.e.
\begin{equation}
\frac{d^{2}}{da^{2}}\hat{f}(a)=V'[\hat{f}(a)]\;.
\end{equation}
This means that we can take $\hat{f}(a)$ to be equal to
$\phi_{cl}(a)$, i.e. $\hat{f}(a)  = \phi_{cl}(a)$ and adjusting
its boundary conditions by an appropriate choice for the value
of the constant $C$ in eq.\,(\ref{first}).

With the above considerations, it is now possible to express the
crossed channel form factor through the variable transformation
$\theta\rightarrow i\pi-\theta$:
\begin{equation}\label{f2}
F_{2}(\theta) \,\equiv\, <0|\,\phi(0)|\,p_{1},\bar{p}_{2}> =
f(i\pi-\theta)\;.
\end{equation}
The analysis of this quantity in infinite volume allows us, in
particular, to get information about the spectrum of the theory.
Its dynamical poles, in fact, located at $\theta^{*}=i(\pi-u)$
with $0<u<\pi$, coincide with the poles of the kink--antikink
$S$-matrix \cite{ffstructure}, and the relative bound states
masses can be then expressed as
\begin{equation}
m_{(b)} \,= \, 2 M \sin\frac{u}{2}\;.
\end{equation}
It is worth to note that this procedure for extracting the semiclassical
bound states masses is remarkably simpler than the standard DHN method
of quantizing the corresponding classical backgrounds, because in general
these solutions depend also on time and have a much more complicated
structure than the kink ones. Moreover, in non--integrable theories these
backgrounds could even not exist as exact solutions of the field
equations: this happens for example in the $\phi^{4}$ theory, where the
DHN quantization has been performed on some approximate backgrounds
\cite{dashen}.

Once the matrix elements (\ref{f2}) are known, one can estimate
the leading behaviour in $\lambda$ of the spectral function in a
regime of the momenta dictated by our assumption of small kink
rapidity. In infinite volume the spectral function $\rho(p^{2})$
is defined as
\begin{equation}
<0|\,\phi(x)\phi(0)|\,0>\equiv\int
\frac{d^{2}p}{(2\pi)^{2}}\,\rho(p^{2})\,e^{i p\cdot x}\;,
\end{equation}
and has the form factor expansion
\begin{equation}
\rho(p^{2})=2\pi\sum\limits_{n}\frac{1}{n!}\int
d\Omega_{1}...d\Omega_{n}\,\delta(p^{0}-E_{1}...-E_{n})\,
\delta(p^{1}-p_{1}...-p_{n})|<0|\,\phi(0)|\,n>|\,^{2}\;,
\end{equation}
with $d\Omega\equiv\frac{d p}{2\pi\,2 E}=\frac{d\theta}{4\pi}$.
The delta functions in the above expression make meaningful the
use of our form factors, derived in the small $\theta$
approximation, only if we consider a regime in which $p^{0}\simeq
M$ and $p^{1}\ll M$ and, from now on, we will always understand this
restriction. The leading $O(1/\lambda)$ contribution\footnote{About
the orders in $\lambda$ of the various form factors, we refer to
the complete discussion in \cite{raj,goldstone}. Our formalism is
slightly different because the $M$ factor in front of (\ref{ffinf}),
which is a natural consequence of considering the rapidity as the basic
variable, increases by $1/\lambda$ the order of all form factors
in the kink sector with respect to \cite{goldstone}; however,
since $\theta\simeq O(\lambda)$, in the final expression for the
spectral function all orders recombine consistently.} to the spectral
function, denoted in the following by $\hat\rho(p^2)$, is given by
the trivial vacuum term plus the kink-antikink contribution:
\begin{equation}\label{rhoinf}
\hat\rho(p^{2}) =
2\pi\delta(p^{0})\delta(p^{1})|<0|\,\phi(0)|\,0>|\,^{2} +
\frac{\pi}{4}\,\frac{\delta\left(\frac{p^{0}}{M}-2\right)}{ M^{2} }\int
\frac{d\theta_{1}}{2\pi}
\left|F_{2}\left(2\theta_{1}+i\pi-\frac{p^{1}}{M}\right)\right|^{2}\;,
\end{equation}
where the range of integration of the above quantity is of order $p^{1}/M$ (note
that, being $p^{1}/M\ll 1$, the integral can be roughly estimated
by evaluating $|F_{2}|^{2}$ at $\theta_{1}=0$: this is what we
will do in the next Sections).

\vspace{0.5cm}

The application of this procedure to the finite volume
case is straightforward, thanks to the possibility of choosing
$\hat{f}(a)$ as a solution of eq.\,(\ref{first}) with any constant
$C$. As explicitly shown by the examples discussed in the next sections,
this is equivalent to define a kink solution configuration on a finite
volume, with the constant $C$ directly related to the size of the system.
We have now to consider the matrix elements of $\phi(0)$ between two
eigenstates $|p_{n_{1}}>$ and $|p_{n_{2}}>$ of the finite volume hamiltonian.
These states can be naturally labelled with the so-called "quasi-momentum"
variable $p_{n}$, which corresponds to the eigenvalues of the translation
operator on the cylinder (multiples of $\pi/L$), and appears in the space
dependent part of eq.\,(\ref{heis}) in the case of finite volume. The TBA
equations \cite{ZamBethe}, valid for large $L$, are exactly a
relation between this variable and the free momentum $p^{\infty}$
of the infinite volume asymptotic states, through a phase shift
$\delta(p^{\infty})$ which encodes the information about the
interaction:
\begin{equation}
p_{n}^{\infty}+\delta(p_{n}^{\infty})=\frac{2n\pi}{L}\equiv
p_{n}\;.
\end{equation}
Defining $\theta_{n}$ as the "quasi-rapidity" of the kink states
by
\begin{equation}
p_{n}=M(L)\sinh\theta_{n}\simeq M(L)\theta_{n}\;,
\end{equation}
we can now write the form factor at a finite volume by replacing
the Fourier integral transform with a Fourier series expansion:
\begin{equation}\label{ff}
f(\theta_{n})\equiv M(L)\int\limits_{-L/2}^{L/2}
da\,e^{i\,M(L)\theta_{n} a}\hat{f}(a)\;,
\end{equation}
\begin{equation}\label{inverseff}
\hat{f}(a)\equiv\frac{1}{L\,M(L)}
\sum\limits_{n=-\infty}^{\infty}\,e^{-i\,M(L)\theta_{n}
a}f(\theta_{n})\;,
\end{equation}
where
\begin{equation}
M(L)\theta_{n}\simeq
p_{n_{1}}-p_{n_{2}}=\frac{(2n_{1}-1)\pi}{L}
- \frac{(2n_{2}-1)\pi}{L}\equiv\frac{2n\pi}{L}\;.
\end{equation}
Since the energy eigenvalues of the finite volume hamiltonian cannot be related
to the quasi-rapidity as in eq.\,(\ref{rapidity}), in principle we are not
allowed to express the crossed channel form factor $F_{2}(\theta)$ via the
change of variable $\theta\rightarrow i\pi-\theta$. However, it is easy
to show that in our regime of approximations the deviations of the kink energy
from (\ref{rapidity}) are of higher order in the coupling and can be neglected
at this stage.

On the cylinder, the spectral function can be expressed as
a series expansion on the form factors:
\begin{equation}\label{rhocyl}
\rho(E_{k},p_{k})
\,=\,
2\pi\sum\limits_{n}\frac{1}{n!}\frac{1}{(2L)^{n}}\sum\limits_{k_{1},...,k_{n}}
\frac{1}{E_{k_{1}}E_{k_{2}}...E_{k_{n}}}\,
\delta(E_{k}-E_{k_{1}}...-E_{k_{n}})\,\delta(p_{k}-p_{k_{1}}...-p_{k_{n}})
\times \end{equation}
\begin{equation*}
\times|<0|\,\phi(0)|\,n>|\,^{2}\;,
\end{equation*}
where $p_{k_{i}}$ are the quasi-momenta of the intermediate states, and
$E_{k_{i}}$ are the finite volume energy eigenvalues, to be determined by
other means (see the comment in Sect.\ref{intro}). In our semiclassical regime,
however, this is not necessary: in fact, in order to evaluate the leading
contribution $\hat\rho(E_{k},p_{k})$, we can consistently approximate the kink
energies with their classical values (of order $1/\lambda$), which
can be exactly computed as a function of the volume. We then have
\begin{equation}
\hat\rho(E_{k},p_{k})
\,= \, 2\pi\delta(E_{k})\delta(p_{k})|<0|\,\phi(0)|\,0>|\,^{2} + \frac{\pi}{4}\,
\frac{\delta\left(\frac{E_{k}}{M}-2\right)}
{M^{2}}\sum\limits_{\theta_{k_{1}}}\left|F_{2}\left(2\theta_{k_{1}}
+ i\pi-\frac{p_{k}}{M}\right)\right|^{2}\,\,\,.
\end{equation}
As in the infinite volume case, the consistency of the semi--classical
approximation selects as the relevant values of the above series those
with $\theta_k \simeq 0$ and therefore it can can be roughly estimated by
simply evaluating $|F_{2}|^{2}$ at $\theta_{k_{1}}=0$.

\vspace{3mm}

It is now interesting to apply these general considerations to the
analysis of the form factors of the fundamental field $\phi(x)$
both for an integrable and a non--integrable QFT.

\section{Sine-Gordon model}\label{SGsection}

\subsection{Infinite volume}

The Sine-Gordon model, defined in (\ref{SGpot}), is an integrable
quantum field theory, for which the infinite volume form factors are
exactly known \cite{ffstructure}. The (anti)soliton background is
given by
\begin{equation}\label{SGkinkinf}
\phi_{cl}(x)=\frac{4}{\beta}\arctan(e^{\pm mx})\;.
\end{equation}
Its classical energy is ${\cal E}_{cl} = 8\frac{m}{\beta^{2}}$, and the first
quantum corrections are known to be of higher order in $\beta^{2}$ \cite{dashen}.
Hence we can consistently approximate the mass $M_{\infty}$ with
this value and assume that the (anti)soliton rapidity will be of
order $\beta^{2}$. The semi--classical form factor (\ref{ffinf})
is explicitly given by
\begin{equation}
f(\theta)=\frac{4
M_{\infty}}{\beta}\int\limits_{-\infty}^{\infty}da \, e^{i( M_{\infty}
\theta)\, a}\,\arctan\left(e^{m a }\right)=\frac{2\pi
i}{\beta}\,\frac{1}{\theta\,\cosh\left[\frac{4\pi}{\beta^{2}}\,
\theta\right]}+\frac{2\pi^{2}}{\beta^{3}}
\delta\left(\frac{\theta}{\beta^{2}}\right)\;.
\end{equation}
It is an interesting check to see whether our formulation of the
semi--classical form factor in terms of the rapidity variable,
i.e. the expression $F_2(\theta)$, matches with the
semi--classical limit of the exact one. In doing this check, the
only thing to take into account is that, in the definition of the
exact form factor of the fundamental field $\phi(x)$, the
asymptotic two--particle state is actually given by the
antisymmetric combination of soliton and antisoliton. Since at our
level the form factor between antisoliton states is simply
\begin{equation}
<\bar{p}_{1}|\,\phi(0)|\,\bar{p}_{2}>=\frac{4
M_{\infty}}{\beta}\int\limits_{-\infty}^{\infty}da\,
e^{i (M_{\infty} \theta) \, a}\,\arctan\left(e^{-m a }\right) =
f(-\theta)\;,
\end{equation}
we finally obtain
\begin{equation}
F_{2}(\theta)=\frac{4\pi
i}{\beta}\,\frac{1}{(i\pi-\theta)\,
\cosh\left[\frac{4\pi}{\beta^{2}}\,(i\pi-\theta)\right]}\;,
\end{equation}
which indeed coincides with the exact result in the regime
$i\pi-\theta\simeq O(\beta^{2})$ \cite{ffstructure}. Furthermore, we
can also check that the dynamical poles of this quantity, located at
\begin{equation}
\theta_{n}=i\pi\left[1-\frac{\beta^{2}}{8\pi}(2n+1)\right]\,,\quad\qquad
-\frac{1}{2}<n<-\frac{1}{2}+\frac{4\pi}{\beta^{2}} \,,
\end{equation}
consistently reproduce the odd part of the well-known
semiclassical breathers spectrum \cite{dashen}
\begin{equation}
m_{b}^{(2n+1)}
=
2M_{\infty}\sin\left[\frac{\beta^{2}}{16}(2n+1)\right]
= (2n+1)\,m\left[1-\frac{(2n+1)^{2}}{3\times 8^{3} }\beta^{4}+...\right]
\end{equation}
Since in the vacuum sector $<0|\,\phi|\,0>=0$, in this model
the $1/\beta^{2}$ leading contribution to the spectral function
takes the form:
\begin{equation}
\hat\rho(p^{2}) \,=
\,4\pi^{3}\,\delta\left(\frac{p^{0}}{M}-2\right)
\frac{1}{\beta^{2}(p^{1})^{2}\,\cosh^{2}
\left[\frac{\pi}{2}\,\frac{p^{1}}{m}\right]}\;.
\end{equation}

\subsection{Finite volume}

\begin{figure}[h]
\hspace{3cm} \psfig{figure=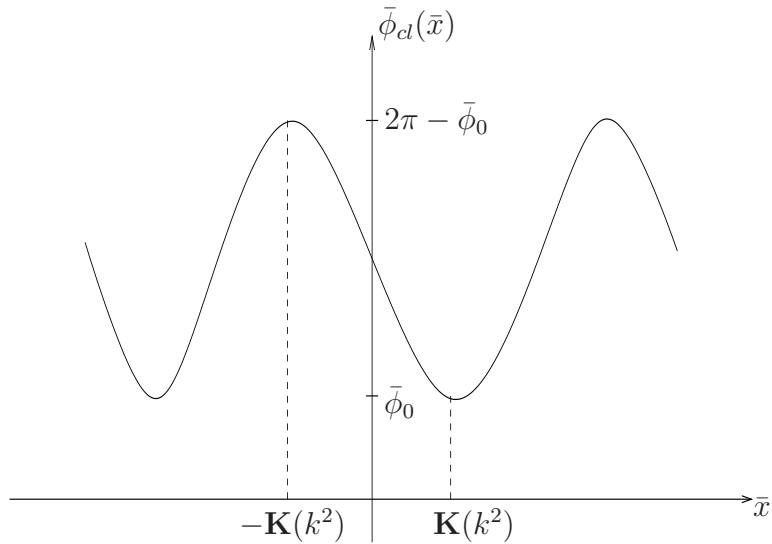,height=7cm,width=10cm}
\vspace{0.1cm} \caption{Sine-Gordon antisoliton in finite volume}
\label{figSGkink}
\end{figure}

In order to consider the effects of a cylindrical geometry, let's
integrate eq.\,(\ref{first}) with a nonzero constant $C$. Rescaling
for convenience the variables as
\begin{equation}\label{scaledvar}
\bar{\phi}=\beta\phi\,,\qquad \bar{x}=m x\,,\qquad
\bar{C}=\frac{\beta^{2}}{m^{2}}C\;,
\end{equation}
a solution of eq.\,(\ref{first}), with $-2<\bar{C}<0$, can be
expressed as
\begin{equation}\label{SGkink}
\bar{\phi}_{cl}(\bar{x}) =
2\arccos\left[\sqrt{\frac{\bar{C}+2}{2}}\;
\textrm{sn}(\pm\bar{x},k^{2})\right]\;,
\end{equation}
where $\textrm{sn}(\bar{x},k^{2})$ is the Jacobi elliptic function
with modulus $k^{2}=\frac{\bar{C}+2}{2}$ (for its properties, see
\cite{GRA}). The plot of this function is drawn in
Fig.\ref{figSGkink}. For a given value of $\bar{C}$, this solution
oscillates with a period $4 \textbf{K}(k^{2})$ between
$\bar{\phi}_{0}$ and $2\pi-\bar{\phi}_{0}$, where $\bar{\phi}_{0}$
is defined by the condition $V(\bar{\phi}_{0})=-\bar{C}$, and
$\textbf{K}(k^{2})$ is the complete elliptic integral of the first
kind.

\vspace{0.5cm}

\psfrag{phicl(x)}{$\bar{\phi}_{cl}(\bar{x})$}
\psfrag{phi0}{$\bar{\phi}_{0}$} \psfrag{2
pi-phi0}{$2\pi-\bar{\phi}_{0}$}
\psfrag{K(k^2)}{$\textbf{K}(k^{2})$}
\psfrag{-K(k^2)}{$-\textbf{K}(k^{2})$} \psfrag{x}{$\bar{x}$}

\vspace{0.5cm}

The solution (\ref{SGkink}) has been proposed in
Ref.\,\cite{takoka} as a model of a crystal of solitons and
antisolitons in the sine-Gordon theory in infinite
volume\footnote{It is worth mentioning the existence of some
impressive old papers \cite{german} where the basic ideas discussed in
\cite{takoka} were already present.}. In our finite volume case,
the solution (\ref{SGkink}) has to be interpreted, instead, as a
single (anti)soliton defined on a cylinder of circumference
\begin{equation}\label{size}
L=\frac{1}{m}\,2 \textbf{K}\left(\frac{\bar{C}+2}{2}\right)\;.
\end{equation}
Within this interpretation, the periodic oscillations of the solution
represent the soliton circling around the cylinder. Eq.\,(\ref{size})
is the explicit relation between the size of the system and the integration
constant $\bar{C}$; one can consistently recover the infinite volume limit for
$\bar{C}\rightarrow 0$: in this limit $L$ goes to infinity and the
function (\ref{SGkink}) goes to the standard (anti)soliton solution
(\ref{SGkinkinf}).

A strong motivation for the choice of this function as the
(anti)soliton background on the cylinder can be obtained by
analyzing the so-called "classical energy per kink" computed in
\cite{takoka}:
\begin{equation}\label{SGclassen}
{\cal E}_{cl}(L) = \int\limits_{-L/2}^{L/2}dx\left[\frac{1}{2}
\left(\frac{\partial \phi_{cl} }{\partial x}\right)^{2} +
\frac{m^{2}}{\beta^{2}}(1-\cos\beta\phi_{cl})\right] =
8\frac{m}{\beta^{2}}\left[
\textbf{E}(k^{2})-\frac{1}{2}(1-k^{2})\textbf{K}(k^{2})\right]\;,
\end{equation}
where $\textbf{E}(k^{2})$ is the complete elliptic integral of the
second kind. In the $L\rightarrow\infty$ limit (which corresponds
to $k'\rightarrow 0$, with $(k')^{2}\equiv
1-k^{2}=-\frac{\bar{C}}{2}$), ${\cal E}_{cl}(L)$ approaches
exponentially the value $M_{\infty}$; the expansions of
$\textbf{E}$ and $\textbf{K}$ for small $k'$ \cite{GRA}, compared
with
$$
e^{-m L}=\frac{1}{16}(k')^{2}+\cdots\;,
$$
lead to the following expansion for large $L$ of the classical
energy
\begin{equation}\label{SGclassenexp}
{\cal E}_{cl}(L) = M_{\infty} - 32\frac{m}{\beta^{2}} e^{-m L} +
O\left(e^{-2 m L}\right)\;.
\end{equation}
By using the theory of finite volume corrections discussed in
\cite{luscher,klassen}, the behaviour (\ref{SGclassenexp}) can be
put in correspondence with the scattering data of the QFT in
infinite volume\footnote{
In eq.\,(\ref{lusch}), we have only written the term that
is relevant in our semiclassical limit, in which the leading contribution
to the mass is given by ${\cal E}_{cl}(L)$.}:
\begin{equation}\label{lusch}
M(L)-M_{\infty}=-\frac{1}{8 m_{b} M_{\infty}^{2}}\,g_{k
\bar{k}b}^{2}\;e^{-m_{b} L} + \cdots \;,
\end{equation}
where $g_{k\bar{k}b}$ is the 3-particle on-shell coupling of kink,
antikink and elementary boson, whose mass in infinite volume is
$m_{b}=m$. Using the known expression for the Sine-Gordon kink-breather
$S$-matrix \cite{zamzam} and evaluating the limit $\beta\rightarrow 0$ of its
residue, we exactly get the coefficient reported in (\ref{SGclassenexp}).

We can now write the finite volume form factor (\ref{ff}) in terms
of the antisoliton background (\ref{SGkink}):
\begin{equation}\label{SGff}
f(-\theta_{n})=\frac{2 M }{\beta}\int
\limits_{-L/2}^{L/2}da\,e^{i\,M\theta_{n}
a}\arccos\left[\sqrt{\frac{\bar{C}+2}{2}}\; \textrm{sn}(m
a)\right]=
\end{equation}
\begin{equation*}
=-\frac{2}{\beta\theta_{n}}\left[e^{iM\theta_{n}\frac{L}{2}}
\log\left(k+i k'\right)-
e^{-iM\theta_{n}\frac{L}{2}}\log\left(-k+i k'\right)
\right]-\frac{2\pi i}{\beta}\,\frac{1}{\theta_{n}\,
\cosh\left[\frac{\textbf{K}'}{m}\,M\theta_{n}\right]}\;,
\end{equation*}
where $k'=\sqrt{1-k^{2}}$ and
$\textbf{K}'(k^{2})=\textbf{K}(k'^{2})$. In order to obtain this
result one has to use the relation
\begin{equation}
\arccos\left[k\, \textrm{sn}(m a)\right]=\frac{1}{i}\log\left[k\,
\textrm{sn}(m a)+i\,\textrm{dn}(m a)\right]\;,
\end{equation}
and, after an integration by parts, finally compare the inverse Fourier
transform (\ref{inverseff}) with the expansion \cite{GRA}
\begin{equation}
\textrm{cn}(m a)
=
\frac{2\pi}{k}\,\frac{1}{mL}\sum\limits_{n=1}^{\infty}
\frac{\cos\left[\frac{(2n-1)\pi}{L}a\right]}
{\cosh\left[\frac{(2n-1)\pi}{L}\,\frac{\textbf{K}'}{m}\right]}\;.
\end{equation}

The form factor (\ref{SGff}) has the correct IR limit\footnote{The
function $\frac{e^{-ixL/2}}{x}$ can be shown to tend to
$-i\pi\delta(x)$ in the distributional sense for
$L\rightarrow\infty$, and in the same way one can show that
$\frac{\cos(xL/2)}{x}$ tends to zero.}, and leads to the following
expressions for $F_{2}(\theta)$ and for the spectral function:
\begin{equation}\label{SGfinvolf2}
F_{2}(\theta_{n})=\frac{4\pi i}{\beta(i\pi-\theta_{n})}\,
\left\{\frac{1}{\cosh\left[\frac{M}{m}\textbf{K}'\,(i\pi-\theta_{n})\right]}+
\left(-1+\frac{2}{\pi}\arctan\frac{k'}{k}\right)\,
\cos\left[M(i\pi-\theta_{n})\frac{L}{2}\right] \right\}\;,
\end{equation}
\begin{equation}\label{SGfinvolrho}
\hat\rho(E_{n},p_{n}) =
4\pi^{3}\,\delta\left(\frac{E_{n}}{M}-2\right)\frac{1}{\beta^{2}(p_{n})^{2}}
\left\{\frac{1}{\cosh\left[\frac{\textbf{K}'}{m}\,p_{n}\right]}+
\left(-1+\frac{2}{\pi}\arctan\frac{k'}{k}\right)\,\cos\left[p_{n}
\frac{L}{2}\right] \right\}^{2}\;.
\end{equation}
Note that the finite volume dependence of both the form factor
(\ref{SGfinvolf2}) and the spectral function (\ref{SGfinvolrho})
is not restricted to the second term only. The
$M(L)\textbf{K}'(k^{2})$ factor in the first term carries the main
$L$-dependence, although it is not manifest but implicitly defined
by eq.\,(\ref{size}).

\section{$\phi^{4}$ field theory in the broken symmetry
phase}\label{phi4section}

The semiclassical analysis performed on the Sine-Gordon
model can be repeated very similarly for the quantum field theory
defined by the potential (\ref{phi4pot}). There is, however, the
important conceptual difference that this QFT is non--integrable:
as we are going to show, this gives us the possibility of estimating
quantities of this quantum field theory which were unknown even in
the infinite volume case.

\subsection{Infinite volume}\label{phi4sectioninfvol}

The standard (anti)kink background is given in this case by
\begin{equation}\label{phi4kinkinf}
\phi_{cl}(x) \, = \,(\pm) \,\frac{m}{\sqrt{\lambda}}
\tanh\left(\frac{m x}{\sqrt{2}}\right)\;,
\end{equation}
with classical energy
$M_{\infty}=\frac{2\sqrt{2}}{3}\frac{m^{3}}{\lambda}$.

With our formulation in terms of the rapidity, we can write an
unambiguous\footnote{This has to be contrasted with the tentative
covariant extrapolation discussed in \cite{goldstone}.}
Lorentz covariant expression for the form factor (\ref{ffinf})
\begin{eqnarray}
<p_{2}|\,\phi(0)|\,p_{1}> & = & M_{\infty}\frac{
m}{\sqrt{\lambda}}\int\limits_{-\infty}^{\infty}da\,e^{iM_{\infty}\theta
a}\,\tanh\left(\frac{m a}{\sqrt{2}}\right) = \nonumber \\
&& = \frac{4}{3}i\pi
\left(\frac{m}{\sqrt{\lambda}}\right)^{3}
\frac{1}{\sinh\left(\frac{2}{3}\pi\frac{m^{2}}{\lambda}\,\theta\right)}\;.
\label{phi4ffinfvol}
\end{eqnarray}
It is of great interest to analyze in this case the dynamical
poles of $F_{2}(\theta)$ for extracting information about the spectrum
of this theory. They are located at
\begin{equation}
\theta_{n}=i\pi\left[1-\frac{3}{2\pi}\,\frac{\lambda}{m^{2}}\,n\right]\,,
\quad\qquad
0<n<\frac{2\pi}{3}\frac{m^{2}}{\lambda} \,,
\end{equation}
and the corresponding bound states masses are given by
\begin{equation}\label{phi4boundst}
m_{b}^{(n)} = 2M_{\infty}\sin\left[\frac{3}{4}\,
\frac{\lambda}{m^{2}}\,n\right] =
n\,\sqrt{2}\,m\left[1-\frac{3}{32}\,
\frac{\lambda^{2}}{m^{4}}\,n^{2}+...\right]\,.
\end{equation}
Note that the leading term is consistently given by multiples of
$\sqrt{2}m$, which is the known mass of the elementary boson.
Contrary to the Sine-Gordon model, we now have all integer
multiples of this mass and not only the odd ones: this is because
we are in the broken phase of the theory, where the invariance
under $\phi\rightarrow -\phi$ is lost. Furthermore, this spectrum
exactly coincides with the one derived in \cite{dashen} by
building approximate classical solutions to represent the
"breathers".

Another important information can be extracted from the residue of
$F_{2}(\theta)$ on the pole corresponding to the lightest bound
state $b^{(1)}$. This quantity, indeed, has to be proportional to
the one-particle form factor $<0|\,\phi|\,b^{(1)}>$  through the
semiclassical 3-particle on-shell coupling of kink, antikink and
elementary boson $g_{k\bar{k}b}$, a quantity so far unknown in this non
integrable theory:
\begin{equation}
\textrm{Res}\,_{\theta=\theta_{1}}F_{2}(\theta)
= i\,\frac{g_{k\bar{k}b}}{2\sqrt{2}M_{\infty}\,
m_{b}^{(1)} }\, <0|\,\phi|\,b^{(1)}> \,\,\,.
\end{equation}
Since the one-particle form factor takes the constant value
$1/\sqrt{2}$, at leading order in the coupling we get
\begin{equation}\label{gphi4}
g_{k\bar{k}b}=\frac{32\, m^{5}}{3\,\lambda^{3/2}}\;.
\end{equation}
Finally, the $1/\lambda$ leading contribution of the
spectral function is given in this case by
\begin{equation}
\hat\rho(p^{2}) =
\frac{2\pi}{\lambda}\,\delta\left(p^{0}/m\right)\delta\left(p^{1}/m\right)
+\frac{\pi^{3}}{2\lambda}\,\delta\left(\frac{p^{0}}{M}-2\right)\frac{1}{
\sinh^{2}\left[\frac{\pi}{\sqrt{2}}\,\frac{p^{1}}{m}\right]}\;.
\end{equation}

\subsection{Finite volume}

Our proposal to describe the finite volume (anti)kink is to use
the following solution of the differential equation (\ref{first})
\begin{equation}\label{phi4kink}
\bar{\phi}_{cl}(\bar{x}) \,=\,(\pm) \,
\sqrt{2-\bar{\phi}_{0}^{2}}\,\;
\textrm{sn}\left(\frac{\bar{\phi}_{0}}{\sqrt{2}}\,
\bar{x},k^{2}\right)\;,
\end{equation}
with $k^{2} = \frac{2}{\bar{\phi}_{0}^{2}}-1$, $V(\phi_{0}) = -C$ and
$1 <\bar{\phi}_{0} < \sqrt{2}$, where we have rescaled the variables as
\begin{equation}\label{scaledvarphi4}
\bar{\phi}\,=\, \frac{\sqrt{\lambda}}{m}\,\phi\,,\qquad
\bar{x}\, =\, m x\;.
\end{equation}
This function oscillates with period
$$
2 \bar{L} \,=\, 4 \frac{\sqrt{2}}{\bar{\phi}_{0}}\,
\textbf{K}(k^{2})
$$
between the two values $-\sqrt{2-\bar{\phi}_{0}^{2}}\,$ and
$\sqrt{2-\bar{\phi}_{0}^{2}}$, and satisfies the antiperiodic
boundary condition $\phi(x+L)=-\phi(x)$. Moreover, it goes to
the standard (anti)kink solution (\ref{phi4kinkinf}) for
$\bar{\phi}_{0}\rightarrow 1$, i.e. when $\bar{L}\,
\rightarrow \, \infty$.

For the "classical energy per kink" in this case we find
\begin{equation}\label{phi4classen}
{\cal E}_{cl}(L)
\, = \, \frac{m^{3}}{\lambda}\,
\frac{\sqrt{2}}{\bar{\phi}_{0}}
\left(-\frac{1}{6}\bar{\phi}_{0}^{4} \, \textbf{K}(k^{2})
+ \frac{1}{3}\bar{\phi}_{0}^{2} \,\left[2 \textbf{E}(k^{2})
- \textbf{K}(k^{2})\right] + \frac{\textbf{K}(k^{2})}{2}\right)
\,\,\,,
\end{equation}
which for $L\rightarrow\infty$ indeed reproduces $M_{\infty}$.
Taking into account the $k\rightarrow 1$ ($k'\rightarrow 0$)
expansions of $\textbf{E}$ and $\textbf{K}$ \cite{GRA} and noting
that
$$
e^{-\sqrt{2}m L}=\frac{1}{256}(k')^{4}+\cdots\;,
$$
we derive the following asymptotic expansion of ${\cal E}_{cl}$
for large $L$:
\begin{equation}
{\cal E}_{cl}(L)=M_{\infty}-8\sqrt{2}\,\frac{m^{3}}{\lambda}
e^{-\sqrt{2}m L}+O\left(e^{-2 \sqrt{2}m L}\right)\;,
\end{equation}
(note that in this theory the mass of the elementary boson is
$m_{b}=\sqrt{2}m$). By using eq.\,(\ref{lusch}), it is easy to see
that this expansion exactly reproduces the value (\ref{gphi4}) for
the 3-particle coupling $g_{k\bar{k}b}$, previously obtained by looking
at the residue of the form factor in infinite volume.

Comparing the inverse Fourier transform (\ref{inverseff}) with the
expansion \cite{GRA}
\begin{equation}
\textrm{sn}(u)
\,=\, \frac{\pi}{k\textbf{K}} \,
\sum\limits_{n=1}^{\infty}
\frac{\sin\left[\frac{(2n-1)\pi}{2\textbf{K}}u\right]}
{\sinh\left[\frac{(2n-1)\pi}{2\textbf{K}}\,\textbf{K}'\right]}\;,
\end{equation}
we obtain for the form factor in a finite volume the following expression
\begin{eqnarray}
F_{2}(\theta_{n}) & = & M
\frac{m}{\sqrt{\lambda}}\, \sqrt{2-\bar{\phi}_{0}^{2}}\, \int
\limits_{-L/2}^{L/2} da \, e^{i\,M(i\pi-\theta_{n})a}\textrm{sn}
\left(\frac{\bar{\phi}_{0}}{\sqrt{2}}\,m a\right) \, = \, \nonumber \\
& = & i\pi \sqrt{\frac{2}{\lambda}} M \,
\frac{1}{\sinh\left[\frac{\sqrt{2}}{m\bar{\phi}_{0}}
\textbf{K}'M(i\pi-\theta_{n})\right]}\;,
\end{eqnarray}
so that, the $1/\lambda$ leading contribution to the spectral function
is given by
\begin{equation}
\hat\rho(E_{n},p_{n}) \, = \, \frac{2\pi}{\lambda}\,
\delta\left(E_{n}/m\right)\delta\left(p_{n}/m\right) +
\frac{\pi^{3}}{2\lambda}\,\delta\left(\frac{E_{n}}{M}-2\right)\frac{1}{
\sinh^{2}\left[
\frac{\sqrt{2}}{m\bar{\phi}_{0}}\textbf{K}'p_{n}\right]} \;.
\end{equation}
Again, as in the Sine-Gordon case, the finite volume dependence of
these quantities comes from the factor $M(L)\textbf{K}'(k^{2})$,
where $M(L)$ is the kink mass given by eq.\,(\ref{phi4classen}).

\section{Further developments and open problems}

Although the form factors provide an important piece of
information about the quantum theory, a more complete
understanding of the semi--classical behaviour of Sine-Gordon and
$\phi^{4}$ models requires an extension of the DHN procedure to
finite volume. This would permit to analyze the energy levels of
the kink "particles" and to compute the quantum corrections to
their masses. A detailed discussion of these topics will be
presented somewhere else \cite{next}.

Furthermore, in order to systematically analyze the
spectra of the proposed theories on the cylinder, it seems
necessary to find the non--trivial periodic classical solutions in
the vacuum sector and apply the DHN procedure to them. As a matter of
fact, the identification the kink-antikink bound states masses in infinite
volume from the dynamical poles of $F_{2}(\theta)$ is a very
powerful tool, whose application to other theories is a topic interesting
in itself, but cannot be directly implemented on the cylinder.

Finally, an important open problem which also deserves further
investigation is the evaluation of the higher loop corrections to
the semiclassical energies (masses), form factors and Green
functions. For the infinite volume case, relevant loop
calculations around the non perturbative kink solutions have been
performed for the Sine-Gordon and $\phi^{4}$ models in the papers
\cite{twoloops}. Whether one can extend these loop calculations to
the case of finite volume (cylinder geometry), in order to
estimate the loop corrections to the semiclassical scaling
functions, form factors etc., remains to be seen. This is
particularly interesting in the Sine-Gordon case, because it would
permit to understand whether also in finite volume the
semiclassical results can be related to the exact ones by a simple
redefinition of the coupling.

\vspace{1cm}

\begin{flushleft}\large
\textbf{Acknowledgements}
\end{flushleft}

The authors thank G. Delfino for stimulating discussions. G.S.
acknowledges UNESP, FAPESP and SISSA for financial support, which
makes this collaboration possible. G.M. and V.R. are grateful to
IFT - Sao Paulo for the warm hospitality.

\end{document}